\documentclass[oneside,twocolumn]{revtex4}
\usepackage[english]{babel}
\usepackage{lipsum}
\usepackage{graphicx}
\usepackage{amsmath}
\usepackage{comment}
\usepackage{braket}
\usepackage{tikz}
\usepackage{bm}
\usepackage{siunitx}
\usepackage{microtype}

\usetikzlibrary{shapes.misc}

\definecolor{darkred}{RGB}{217,83,25}
\definecolor{darkblue}{RGB}{0,114,189}

\includecomment{maintext}

\newcommand{\sreq}{Schr\"{o}dinger equation}

\begin{document}
\title{Fast state transfer in a $\Lambda$-system: a shortcut-to-adiabaticity approach to robust and resource optimized control}
\author{Henrik Lund Mortensen, Jens Jakob W. H. S\o{}rensen, Klaus M\o{}lmer and Jacob Friis Sherson}
\email{sherson@phys.au.dk}
\affiliation{Department of Physics and Astronomy, Aarhus University, Ny Munkegade 120, 8000 Aarhus C, Denmark}
\date{\today}
\begin{abstract}
We propose an efficient strategy to find optimal control functions for state-to-state quantum control problems. Our procedure first chooses an input state trajectory, that can realize the desired transformation by adiabatic variation of the system Hamiltonian. The shortcut-to-adiabaticity (STA) formalism then provides a control Hamiltonian that realizes the reference trajectory exactly but on a finite time scale. As the final state is achieved with certainty, we define a cost functional that incorporates the resource requirements and a perturbative expression for robustness. We optimize this functional by systematically varying the reference trajectory. We demonstrate the method by application to population transfer in a laser driven three-level $\Lambda$-system, where we find solutions that are fast and robust against perturbations while maintaining a low peak laser power. 

% Shortcut-to-adiabaticity (STA) techniques provides a map between the state trajectory and the corresponding Hamiltonian, defined by a control function. We propose an optimization scheme where a cost function, measuring the quality of the control, is minimized by optimization of the trajectory. We apply the scheme to a population transfer problem in a three-level $\Lambda$-system, where we find solutions that are 60\% faster than the result reported in a recent experiment \cite{du_experimental_2016}, while maintaing the same peak Rabi frequency and robustness against Rabi frequency scaling.
\end{abstract}
\maketitle

\section{Introduction}

% \singlebullet{Control of quantum systems is important}

Precise control of quantum systems is required to realize a number of applications in quantum information and precision quantum measurements \cite{divincenzo,q_comp_and_info,Giovannetti1330}. The simultaneous fulfillment of constraints on protocol duration, fidelity, robustness against parameter variations, or feasibility to implement the desired interaction, can be handled with optimal control theory \cite{pontryagin,tannor,reich_krotov,crab,dcrab}. Here, a cost functional is defined that quantifies the quality of a solution and penalizes, e.g., extreme values or strong variations in the control fields \cite{georg_jaeger,QCT_phd_tutorial}. This cost functional is then numerically optimized. Building on pioneering efforts on selective excitation of molecular systems and NMR \cite{Glaser2015}, optimal control theory has been successfully applied in various quantum control problems, including manipulation of Bose-Einstein condensates \cite{van_frank,vib_state_inv} and transport of single atoms in optical tweezers \cite{sorensen_exploring_2016}. Benign control problems permit the use of gradient methods on a large number of control parameters. However, for time constrained problems, these methods may converge on sub-optimal solutions, and thus require multi-start approaches with different initial guesses. This process is numerically inefficient and offers no guarantee for the identification of the optimal solution \cite{sorensen_exploring_2016}. We recall that the evaluation of a cost function involving the transfer fidelity requires the numerical solution of the \sreq{}, which is often time consuming.

In this paper we propose an extension to a method which by construction is guaranteed to reach the desired final state \cite{berry_transitionless_2009,STA_dressed_states,chen_lewis-riesenfeld_2011}. This implies that numerical solution of the \sreq{} is not necessary. Thus, we need only assess the constraints on duration, resources and robustness of the protocol, which are all explicitly evaluated for each candidate solution. The starting point of the analysis is to define a time dependent trajectory $\phi_0(t)$ for our wave function reaching from the initial $\phi_i$ to the desired final state $\phi_f$. We construct $\phi_0(t)$ as a non-degenerate eigenstate of a Hamiltonian, $H_0(t)$. Subject to a slowly varying $H_0(t)$, the time dependent solution of the \sreq{} will adiabatically follow $\phi_0(t)$  and end up in the final state with certainty \cite{STA_by_CD}. The requirement of adiabaticity, however, only permits solutions of long duration, which may be useless due to the time resource available or due to the effect of decoherence mechanisms. 

\begin{figure*}
\hspace{-1cm}
\input{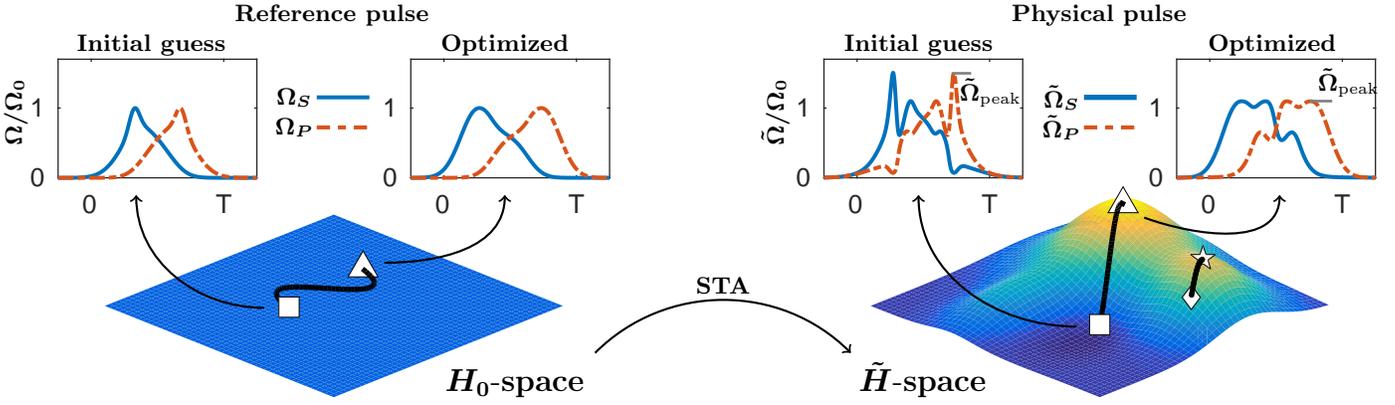}
\caption{Overview of the optimization strategy. An initial guess for the reference Hamiltonian $H_0$, represented by the laser pulse amplitudes $\Omega_{S/P}$, is mapped by STA to the physical Hamiltonian, $H$, and corresponding physical pulses, Eqs.~(\ref{tilde P explicit},\ref{tilde S explicit}). The physical Hamiltonian is then given a score according to the cost functional, Eq. (\ref{cost function}), designed to reflect the experimental constraints. We search through $H_0$-space to find reference pulses which correspond to a local optimum (white triangle) in $\tilde{H}$-space. The details of the optimization process is described in section \ref{parametrization}. The pulses shown result from a single optimization at $T = 0.4$ \si{\milli\second}. The white diamond and white star represent an optimization process, where a Gaussian shape was used as an initial guess for the reference pulses. The optimization reaches a local but not a global optimum illustrating the need to do global optimization by multi-starting with a range of initial guesses.   % For an initial guess (white square), the parameters are locally optimized such that the physical pulses minimize the cost functional EQREF.
}
\label{fig1_H0_and_H_space}
\end{figure*}

To speed up adiabatic protocols, we therefore invoke shortcut-to-adiabaticity (STA) \cite{berry_transitionless_2009,STA_dressed_states,chen_lewis-riesenfeld_2011}. STA constructs an explicitly modified Hamiltonian $\tilde{H}(t)$ which suppresses the non-adiabatic transitions and forces a quantum system to follow the eigenstates of $H_0(t)$, thus maintaining perfect state transfer at finite evolution times. For the purpose of optimization, we now exploit the freedom in choosing the trajectory, \textit{i.e.}, $H_0(t)$, such that the cost functional for the STA modified Hamiltonian is minimized. The method is visualized in Fig. \ref{fig1_H0_and_H_space}, where each point in the $H_0$-space represents a realization of $H_0(t)$, mapped by STA to a time dependent Hamiltonian $\tilde{H}(t)$  shown as a point in the $\tilde{H}$-space. The cost function is indicated as the vertical dimension in $\tilde{H}$-space. We search in $H_0$-space for the trajectory, leading to the physical control Hamiltonian in $\tilde{H}$-space that optimizes the costs. This type of approach has been applied in a two-level system, where the state trajectory was parametrized in order to find optimally robust solutions \cite{ndong_robust_2015}. Previous work has also combined STA and optimal control theory for studying e.g. atom transport \cite{ruschhaupt_optimally_2012, ruschhaupt_shortcuts_2014,chen_optimal_2011,zhang_fast_2015}.

% , and in principle, STA offers arbitrarily fast solutions. However, there is a trade-off between speed and energy requirements.

% In STA there is freedom to construct a Hamiltonian for many state trajectories, not just adiabatic trajectories.

% STA has been applied to realize state-to-state transfers in various systems \cite{kang_fast_2016,du_experimental_2016,zhang_transitionless_2016,STA_transport_2014}.  State transfer between the two lower levels are performed much faster than the adiabatic STIRAP protocol, while still being robust against scaling of the control parameters.

We shall primarily be concerned with the duration and the energy requirements associated with application of strong control fields \cite{Santos2015,campbell_trade-off_2017}. We demonstrate our method by the application to population transfer in a three-level $\Lambda$-system. Recently, Du \emph{et al} \cite{du_experimental_2016} reported a successful experimental application of STA to the STIRAP protocol in this system (see also Ref. \cite{xi_chen_three_level}). By extending the pulse parametrization and optimizing in this space we find solutions which are almost twice as fast as the solution reported in Ref. \cite{du_experimental_2016}, while still satisfying the experimental requirements.

% % maintaining the same level of Rabi frequency peak and robustness (See Fig. \ref{peak_and_q_vs_T}).

%  can be combined with optimization techniques to obtain optimal controls\cite{}.

 % Here, we investigate a three-level system, and use the same pertubative calculation to obtain a measure of sensitivity with respect to variations. In addition to the Rabi frequency peak, we use the sensitivity in a cost functional to be minimized. This way, we find fast solutions, with a compromize between robustness and resource requirements. The method is easily extended to include other cost measures.

 % , transport of atoms in harmonic\cite{} and nonharmonics traps\cite{zhang_fast_2015}.
% In Ref. \cite{ndong_robust_2015}, the authors use inverse engineering in a two-level system, parametrize the trajectory and nullify pertubation terms, such that a robust pulse is derived, which also has low area.

The paper is outlined as follows: In section \ref{Optimization strategy} we explain the optimization strategy in detail. In section \ref{Application} we apply the strategy to the three-level $\Lambda$-system. Finally, in Section \ref{Conclusion} we conclude the paper.

% \singlebullet{ Combine shortcut-to-adiabatic and optimal control to get super good solutions}
% \singlebullet{ Short review of results}
% 'Using shortcut-to-adiabaticity approach, we find protocols which realize the state transfer in half the time compared to the experimental paper \cite{du_experimental_2016} while maintaining robustness.'
% \lipsum[1-2]

% The optimization strategy is based on the freedom

\section{Optimization strategy}
\label{Optimization strategy}
The goal of STA is to transport an initial state, $\phi_i$, to a final state, $\phi_f$ along a chosen state trajectory $\phi_0(t)$ such that $\phi_0(t)$ is the instantaneous eigenstates of a \emph{reference} Hamiltonian, $H_0(t)$, and $\phi_0(0) = \phi_i$ and $\phi_0(T) = \phi_f$. We note that time-evolution with the Hamiltonian $H_0(t)$ does not follow $\phi_0(t)$ due to non-adiabatic transitions among the instantaneous eigenstates. STA suppresses these transitions by introducing a \emph{counter-diabatic} term,
\begin{equation}
  \label{H_cd}
  H_{\text{cd}} = i\sum_n \ket{\partial_t \phi_n}\bra{\phi_n} - \braket{\phi_n|\partial_t \phi_n} \ket{\phi_n}\bra{\phi_n},
\end{equation}
where $\ket{\phi_n(t)}$ are all the instantaneous eigenstates of $H_0(t)$ with eigenvalues $E_n(t)$. Regardless of the process duration, a system subject to the total, physical Hamiltonian,
\begin{equation}
  \label{}
  \tilde{H}(t) = H_0(t) +   H_{\text{cd}}(t) ,
\end{equation}
experiences no transitions among the instantaneous eigenstates. If this Hamiltonian can be realized experimentally, the initial state $\phi_i$ evolves with certainty into $\phi_f$ at time $T$, 
\begin{equation}
  \label{psi}
  \ket{\psi(T)} = e^{i\xi_0(T)}\ket{\phi_0(T)}= e^{i\xi_0(T)}\ket{\phi_f},
\end{equation}
acquiring a dynamical and geometric phase
\begin{equation}
  \label{xi}
  \xi_0(t) = -\int_0^t E_0(t) dt + i \int_0^t \braket{\phi_0(t)|\partial_t \phi_0(t)} dt.
\end{equation}
%This ensures a perfect state transfer at $t = T$.
% Suppose that we search for solutions to a state-to-state transfer problem. We can then choose $H_0(t)$ such that one of the eigenstates coincide with the initial and target state at the beginning and end of the protocol. Realizing the total Hamiltonian then ensures that the system follows this eigenstate and therefore completes the state transfer.
There are many choices for time dependent $H_0(t)$ which solve a given state transfer. We handle the experimental constraints by defining a cost functional on the physical Hamiltonian $\tilde{H}(t)$ and by optimizing $H_0(t)$ such that the cost is minimized (see Fig. \ref{fig1_H0_and_H_space}). 
A unique feature of this optimization strategy compared to traditional optimal control theory, is the guarantee of successful state transfer. Thus, the numerical optimization only focuses on the experimental constraints.

\section{State transfer in a $\Lambda$-system}
\label{Application}
We study application of the optimization strategy to population transfer in a three-level $\Lambda$-system, with two lower states ($\ket{1}$ and $\ket{3}$) and an excited state ($\ket{2}$), which has a very short life time. The two lower levels of such a system can, for example, be used to represent a qubit, while the excited state mediates the coupling between the qubit states \cite{vitanov_stimulated_2016}. We follow Ref. \cite{du_experimental_2016}, where the authors realize the $\Lambda$-system in two hyperfine levels of the ground state and a short-lived optically exited state in $^{87}$Rb atoms. The goal is to transfer population from $\ket{1}$ to $\ket{3}$, while avoiding population in $\ket{2}$ due to its short life time. We assume that a direct coupling is unavailable. Instead, a stimulated Raman transition is used by applying two lasers coupling each of the ground states to the excited state. The adiabatic STIRAP protocol has been successfully applied to this problem \cite{vitanov_stimulated_2017,Stirap.pdf}.

In the rotating wave approximation and under the two-photon resonance condition, the system Hamiltonian is given by ($\hbar = 1$)
\begin{equation}
  \label{STIRAP H}
  H_{\Lambda}(t) = \frac{1}{2}\left( \begin{matrix}
      0 & \Omega_P(t)e^{i\phi_L} & 0 \\
      \Omega_P(t)e^{-i\phi_L} & 2\Delta & \Omega_S(t) \\
     0 & \Omega_S(t) & 0
    \end{matrix} \right),
\end{equation}
where $\Omega_P(t)$ and $\Omega_S(t)$ are the real Rabi frequencies of the pump and Stokes lasers, coupling $\ket{1}$-$\ket{2}$ and $\ket{3}$-$\ket{2}$ respectively. $\Delta$ is the one-photon detuning and $\phi_L$ is the fixed phase difference between the lasers. We consider the $\Lambda$-system under large one-photon detuning, $\Delta \gg \Omega_P(t),\Omega_S(t)$, where the excited state can be adiabatically eliminated, such that the system reduces to an effective two-level system \cite{brion_adiabatic_2007}. Here, the effective Hamiltonian is given by
\begin{equation}
  \label{H eff}
  H_{0}(t) = \frac{-1}{2} \left( \begin{matrix}
      \Delta_{\text{eff}} & \Omega_{\text{eff}} \\
\Omega_{\text{eff}} & -\Delta_{\text{eff}}
    \end{matrix} \right),
\end{equation}
with
\begin{align}
  \label{omega eff}
 \Omega_{\text{eff}} &= \frac{\Omega_P\Omega_S}{2\Delta}, \\
\Delta_{\text{eff}} &= \frac{\Omega_P^2 - \Omega_S^2}{4\Delta}.
  \label{Delta eff}
\end{align}
The eigenstates of the effective Hamiltonian are given by
\begin{align}
  \ket{a_0} &= \cos \theta \ket{1}- \sin\theta \ket{3}  \label{a0}\\
  \ket{a_-} &= \sin \theta \ket{1}+ \cos\theta \ket{3},
  \label{am}
\end{align}
where the mixing angle is defined as $\theta = \arctan \frac{\Omega_P}{\Omega_S}$. Imposing the boundary conditions
\begin{equation}
  \label{boundaries}
  \frac{\Omega_P(0)}{\Omega_S(0)} = 0 \;,\;\;\;\;\frac{\Omega_S(T)}{\Omega_P(T)} = 0,
\end{equation}
ensures that $\ket{a_0(0)} = \ket{1}$ and $\ket{a_0(T)} = -\ket{3}$. The reference pulses $\Omega_P(t)$ and $\Omega_S(t)$ thus define the trajectory $\phi_o(t)$. If the  process duration is large enough, by the adiabatic theorem, the system follows the eigenstate, $\ket{a_0(t)}$, and the STIRAP protocol therefore realizes the state transfer.

\subsection{Shortcut-to-adiabaticity for a STIRAP trajectory}

We can use the STA formalism to follow the eigenstates of $H_0(t)$ even when the system is driven more rapidly. To do this we calculate the counter-diabatic Hamiltonian, Eq. (\ref{H_cd}). The result is
\begin{equation}
  \label{cd H}
  H_{\text{cd}}(t) = \left(\begin{matrix}
      0 & -i\Omega_a \\
    -i\Omega_a & 0
    \end{matrix} \right),
\end{equation}
where $\Omega_a = \frac{\dot{\Omega}_P\Omega_S - \dot{\Omega}_S\Omega_P }{\Omega_P^2 + \Omega_S^2}$.

\begin{figure*}
\begin{tikzpicture}
    \node[anchor=south west,inner sep=0] at (0.1,0.4) {\includegraphics[scale=1]{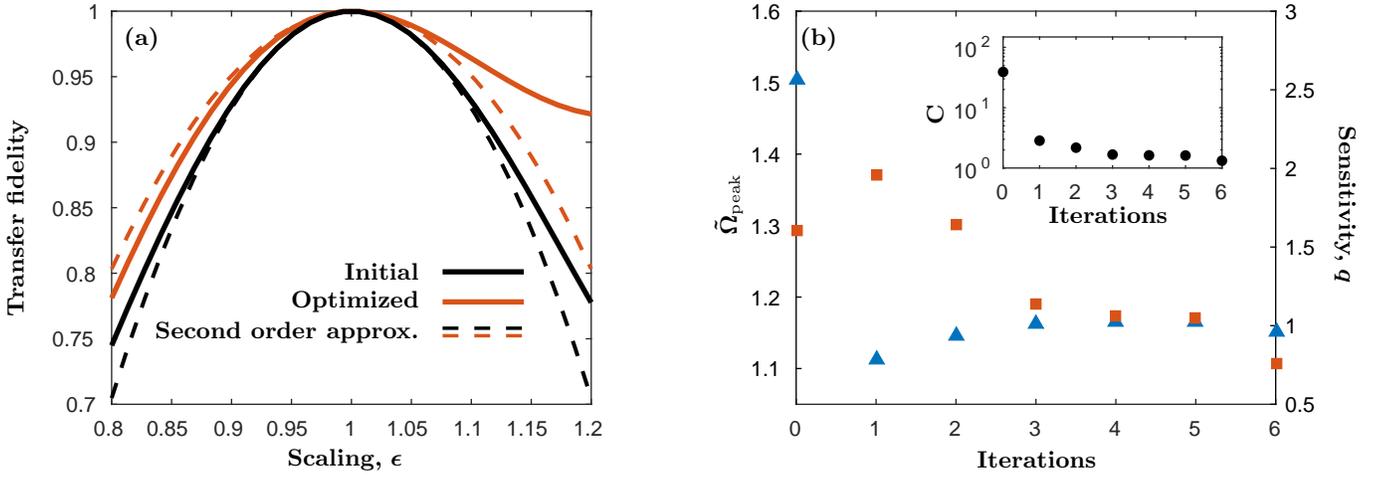}};

    \begin{scope}[shift={(9.4,0)}]
    \node[anchor=center,inner sep=0,rotate=90] at (-0.3,3.5) {\bm{$\tilde\Omega_{\text{peak}}$}};
    \node[anchor=center,inner sep=0,rotate=270] at (7.9,3.5) {\textbf{Sensitivity,} \bm{$q$}};

    \node[anchor=center,inner sep=0] at (3.8,0.1) {\textbf{Iterations}};
    \node[anchor=center,inner sep=0,rotate=90] at (2.45,4.7) {\textbf{C}};
    \node[anchor=south west,inner sep=0] at (3.95,3.24) {\textbf{Iterations}};
    \end{scope}

    % \end{scope}

    % \node[anchor=south west,inner sep=0] at (12.5,3.2) {\textbf{Optimized}};
    % \node[anchor=south west,inner sep=0] at (12.4,6.2) {\textbf{Initial guess}};

    \node[anchor=center] at (1.3,5.7) {\textbf{(a)}};
    \node[anchor=center] at (10.3,5.7) {\textbf{(b)}};

    \begin{scope}[shift={(-10,0)}]
    \def\ysep{0.4};
    \def\linesep{0.2};
    \def\linelength{1.08};
    \def\x{15.1};
    \def\y{2.6}
    \def\dashsep{0.1};
    \def\dashlength{6pt};
    \node[anchor=south west,inner sep=0,rotate=90] at (9.8,2) {\textbf{Transfer fidelity}};
    \node[anchor=center,inner sep=0] at (14.0,0.1) {\textbf{Scaling,} \bm{$\epsilon$}};
      \node[anchor=east] at (\x,\y) {\textbf{Initial}};
      \node[anchor=east] at (\x,\y-\ysep) {\textbf{Optimized}};
      \node[anchor=east] at (\x,\y-\ysep-\ysep) {\textbf{Second order approx.}};

      \draw[black,line width=2] (\x+\linesep,\y) to (\x+\linesep+\linelength,\y);
      \draw[darkred,line width=2] (\x+\linesep,\y-\ysep) to (\x+\linesep+\linelength,\y-\ysep);
      \draw[dash pattern=on \dashlength off \dashlength,black,line width=1.3] (\x+\linesep,\y-\ysep-\ysep+\dashsep/2) to (\x+\linesep+\linelength,\y-\ysep-\ysep+\dashsep/2);
      \draw[dash pattern=on \dashlength off \dashlength,darkred,line width=1.3] (\x+\linesep,\y-\ysep-\ysep-\dashsep/2) to (\x+\linesep+\linelength,\y-\ysep-\ysep-\dashsep/2);
    \end{scope}

    % \draw[color=darkred,line width=2] (\x+\linesep,\y-\ysep) to (\x+\linesep+\linelength,\y\ysep);

    % \node[anchor=south west,inner sep=0] at (2.4,6.2) {\textbf{Optimization progress}};

\end{tikzpicture}
\caption{a) Transfer fidelity shown as a function of the scaling parameter $\epsilon$ on the field amplitudes $\tilde{\Omega}_{P/S} \rightarrow \epsilon\tilde{\Omega}_{P/S}$. The numerical and second order approximation to the fidelity (see Eq. (\ref{fidelity})) is plotted with solid and dashed lines. Optimization reduces the effect of scaling on the transfer fidelity, cf. the difference between the (lower) black and the (upper) red curves. b) Optimization process for $T = 0.4$ \si{\milli\second}. The sensitivity $q$ is represented by the squares and the right axis while the triangles and the left axis represents the peak Rabi frequency $\tilde\Omega_{\text{peak}}$. The cost functional in 
Eq. (\ref{cost function}) (insert) is monotonically decreasing with each iteration, while $\tilde\Omega_{\text{peak}}$ and $q$ settle towards a balanced minimum. }
\label{q_and_iter}
\end{figure*}

Implementing $H(t) = H_{0}(t) + H_{\text{cd}}(t)$ requires temporal control of the relative phase between the Stokes and pump lasers, which can be circumvented by transforming to a frame, defined by the unitary transformation
\begin{equation}
  \label{unitary}
  U = \left( \begin{matrix}
      e^{-i\gamma(t)/2} & 0 \\
     0 & e^{i\gamma(t)/2}
    \end{matrix} \right),
\end{equation}
with $\gamma = \arctan \left( \frac{\Omega_a}{\Omega_{\text{eff}}} \right) + \phi_L$ \cite{du_experimental_2016}. The resulting Hamiltonian
\begin{equation}
  \label{tilde H}
  \tilde{H}(t) = \frac{-1}{2} \left(\begin{matrix}
      \tilde{\Delta}_{\text{eff}} & \tilde\Omega_{\text{eff}} \\
 \tilde\Omega_{\text{eff}} & -\tilde{\Delta}_{\text{eff}}
    \end{matrix} \right),
\end{equation}
can be implemented with a real Rabi frequency $\tilde\Omega_{\text{eff}} = \sqrt{\Omega^2_{\text{eff}} + \Omega^2_a}$ and $\tilde\Delta_{\text{eff}} = \Delta_{\text{eff}} + \dot{\gamma}$, and since the basis state populations are unaffected by this transformation, the new $\tilde{H}(t)$ also realizes the population transfer. We now look for the \emph{physical pulses}, $\tilde\Omega_P(t)$ and $\tilde\Omega_S(t)$, applied to the original three-level system, that realize $\tilde{H}(t)$. We thus solve for the values of $\tilde\Omega_P$ and $\tilde\Omega_S$ that yield Eqs. (\ref{omega eff}) and (\ref{Delta eff}) with $\Omega_{\text{eff}}$, $\Delta_{\text{eff}}$ replaced by $\tilde\Omega_{\text{eff}}$ and $\tilde\Delta_{\text{eff}}$. The result is
% % \begin{align}
% %   \tilde\Delta_{\text{eff}} &= \frac{\tilde\Omega_P^2-\tilde\Omega_S^2}{4\Delta} \\
% %   \tilde\Omega_{\text{eff}} &= \frac{\tilde\Omega_P\tilde\Omega_S}{2\Delta},
% %   \label{tilde vars}
% % \end{align}
% with no need for a temporal control of the phase difference. The result is
\begin{align}
  \tilde\Omega_P(t) &= \sqrt{2\Delta \left(\sqrt{\tilde\Delta_{\text{eff}}^2 +\tilde\Omega_{\text{eff}}^2 } + \tilde\Delta_{\text{eff}} \right)}   \label{tilde P explicit}\\
  \tilde\Omega_S(t) &= \sqrt{2\Delta \left(\sqrt{\tilde\Delta_{\text{eff}}^2 +\tilde\Omega_{\text{eff}}^2 } - \tilde\Delta_{\text{eff}} \right)}.
  \label{tilde S explicit}
\end{align}

 For any choice of the reference pulses, $\Omega_P(t)$ and $\Omega_S(t)$, that fulfill the conditions, Eq. (\ref{boundaries}), we can calculate $\tilde{\Omega}_{\text{eff}}(t)$ and $\tilde{\Delta}_{\text{eff}}(t)$ and therefore the physical pulses $\tilde\Omega_P(t)$ and $\tilde\Omega_S(t)$. 
 As long as the elimination of the excited state remains valid, subjecting the three-level atom to these physical pulses will yield the perfect transfer between the ground states in any finite time interval. Fig. \ref{fig1_H0_and_H_space} illustrates how  
 Eqs. (\ref{tilde P explicit}) and (\ref{tilde S explicit}) represents the mapping between $H_0$-space and $\tilde{H}$-space,

% In the next section we define a cost functional to reflect the experimental constraints on the physical pulses.

\subsection{Cost functional}
For any choice of the reference pulses that fulfill the boundary conditions, STA provides the corresponding physical pulses through Eq. (\ref{tilde P explicit}) and (\ref{tilde S explicit}). However, the physical pulses might violate constraints set by the experiment. The constraints considered here include the peak intensity of the lasers and robustness against a scaling of the control parameters and are based on the experiment reported in Ref. \cite{du_experimental_2016}. 

 % Here, the authors have constraints on the peak intensity of the lasers. Furthermore,

% Any choice of reference pulses, $\Omega_P(t)$ and $\Omega_S(t)$, will lead to perfect population transfer as long as the boundary conditions are fulfilled. However, the corresponding physical pulses, $\tilde\Omega_P(t)$ and $\tilde\Omega_P(t)$, might differ in quality. We measure the quality by a cost functional defined below, which we minimize with a standard optimization algorithm.

The peak intensity of the laser can be included in the cost functional as the dimensionless quantity
\begin{equation}
  \label{peak intensity}
\tilde\Omega_{\text{peak}} =   \max \{ \tilde\Omega_{S/P}(t)\}/\Omega_0,
\end{equation}
where we choose $\Omega_0 = 2\pi \cdot 5$ \si{\mega\hertz} to define the scale. Minimizing $\tilde\Omega_{\text{peak}}$ is equivalent to minimizing the peak intensity since $I_{S/P}  \propto \tilde\Omega_{S/P}^2$.

In experiments with many atoms, the spatial laser profile causes different atoms to experience different laser powers depending on their location. Effectively, this corresponds to a random scaling of the Rabi frequency, $\tilde\Omega \rightarrow \epsilon \tilde\Omega$, where $\epsilon \approx 1$. In the adiabatic limit, this scaling does not alter the state transfer, but the values of the time dependent Rabi frequencies of the physical pulses are important when we apply the STA, and the scaling reduces the transfer fidelity, as shown in Fig. \ref{q_and_iter}. We thus seek solutions which are robust against this perturbation. The sensitivity towards amplitude scaling can be quantified by perturbation theory \cite{ndong_robust_2015}. To second order in $\epsilon - 1$, the correction to the transfer fidelity is found to be
\begin{equation}
  \label{fidelity}
  \mathcal{F} \approx 1 - 4 q \left(\epsilon -1 \right)^2
\end{equation}
where we define the \emph{sensitivity},
\begin{equation}
  \label{q}
  q = \bigg| \int_0^T e^{i \left[\xi_0(t) - \xi_-(t) \right]}\braket{a_{-} (t) |U^\dagger(t) \tilde H(t)U(t) |a_0(t)}dt\bigg|^2,
\end{equation}
where $U(t)$ is given by Eq. (\ref{unitary}) and $\xi_i$ is given by Eq. (\ref{xi}). By minimizing $q$ we minimize the sensitivity for variations in intensity. 

To penalize solutions with large peak Rabi frequency, $\tilde\Omega_{\text{peak}}$, and large values of the sensitivity, $q$, we introduce a cost functional. We define our goal based on the peak value and sensitivity found in Ref. \cite{du_experimental_2016}. Here, two Gaussians are used for the reference pulses at a process duration of $T = 0.4$ \si{\milli\second}, and for such Gaussians we have $\tilde\Omega_{\text{peak}}=1.14$ and $q = 1.59$. We seek to match these values at the lowest possible duration. We heuristically find that a cost functional defined as
\begin{equation}
  \label{cost function}
  C = \exp \left[ 10 \left(\tilde\Omega_{\text{peak}} - 1.14 \right) \right] + \exp \left[ 2 \left(q - 1.59 \right) \right]
\end{equation}
represents a balanced minimum of $\tilde\Omega_{\text{peak}}$ and $q$ in accordance with our goal.

\begin{figure*}
\hspace{-0.7cm}
\begin{tikzpicture}
      \node[anchor=south west,inner sep=0] at (0.1,0.4) {  \includegraphics[scale=1]{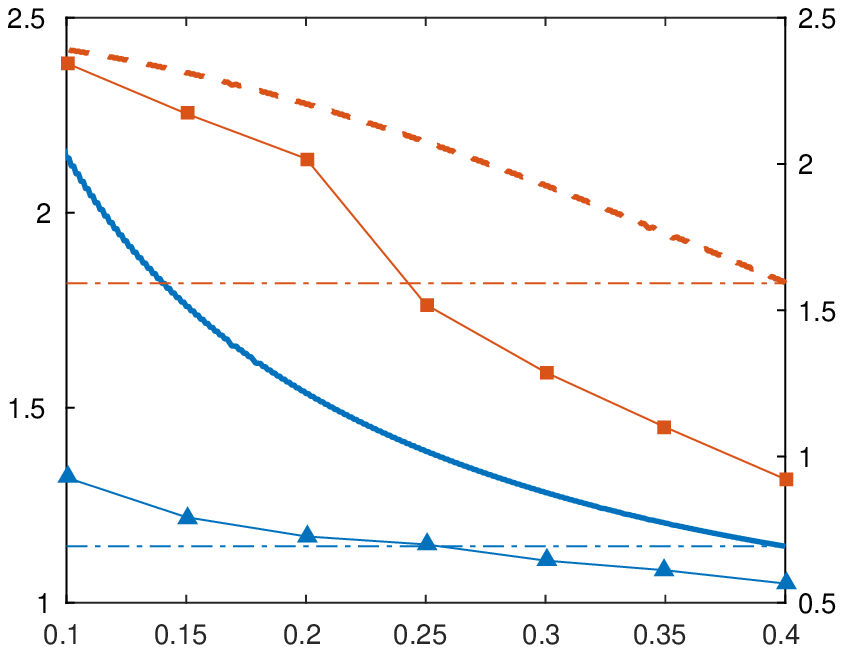}};
      \node[anchor=center,inner sep=0,rotate=90] at (0.5,3.8) {\bm{$\tilde\Omega_{\text{peak}}$}};
      \node[anchor=center,inner sep=0,rotate=-90] at (9.2,3.8) {\textbf{Sensitivity,} \bm{$q$}};
      \node[anchor=center,inner sep=0] at (4.9,0) {\textbf{T} [ms]};
      % \node[anchor=center,inner sep=0] at (4.9,7) {\textbf{Optimization result}};

      \node[anchor=center] at (7.9,6.4) {(a)};

      \def\x{4.675};
      \def\y{3.15};
      \def\r{0.5};
      \def\lwidth{0.6};
      \def\style{dashed};

      \draw[\style,line width=\lwidth] (\x,\y) circle (\r);

      \node[anchor=center] at (\x,\y) {\bm{$T_{\text{\textbf{eqv}}}$}};

      \draw[\style,line width=\lwidth] (\x,\y+\r) to (\x,6.75);
      \draw[\style,line width=\lwidth] (\x,0.85) to (\x,\y-\r);

      \node[anchor=south west,inner sep=0] at (10.4,0.4) {  \includegraphics[scale=1]{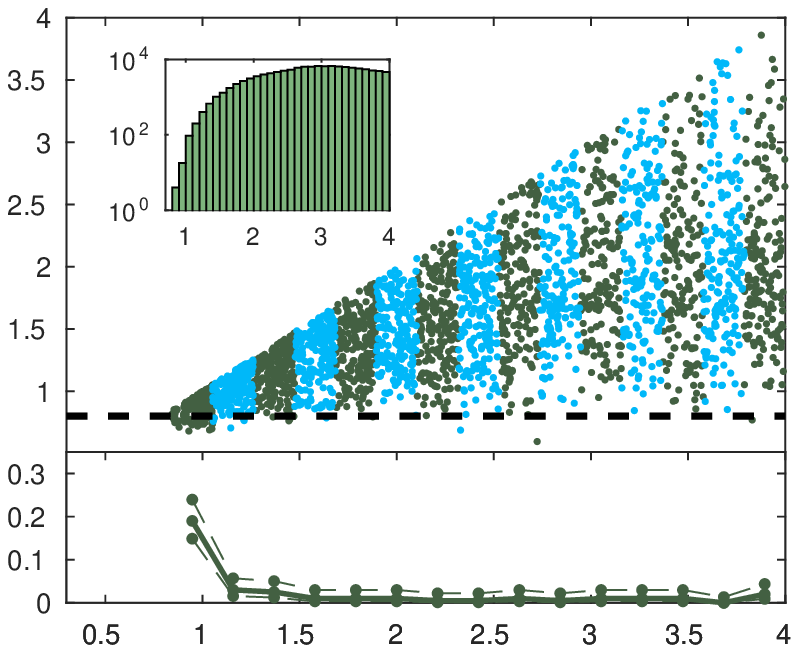}};

      \node[anchor=center] at (14.3,0) {\textbf{log $C$ before optimization}};

      \node[anchor=center] at (7.85+9,6.4) {(b)};

      \def\x{9.8};
      \def\y{1.7};
      \node[anchor=center,rotate=90] at (\x,\y) {\textbf{Probability}};
      \node[anchor=center,rotate=90] at (\x+0.25,\y) {\textbf{for success}};

      \def\x{9.8};
      \def\y{4.7};
      \node[anchor=center,rotate=90] at (\x+0.25/2,\y) {\textbf{log $C$ after optimization}};

      \def\x{13.2};
      \def\y{4.3};

      \node[anchor=center] at (\x,\y) {\scriptsize{\textbf{log $C$ before}}};
      \node[anchor=center] at (\x,\y-0.2) {\scriptsize{\textbf{opt.}}};
      \node[anchor=center,rotate=90] at (\x-1.9,\y+1.2) {\scriptsize{\textbf{Counts}}};

\end{tikzpicture}
\caption{(a) This figure presents the result of optimizing the STA pulse sequences according to the cost functional, Eq. (\ref{cost function}). Red squares show the optimized sensitivity and blue triangles show the optimized peak Rabi frequency. The dashed-dotted lines show the target values of $\tilde\Omega_{\text{peak}} = 1.14$ and $q=1.59$ obtained in Ref. \cite{du_experimental_2016} by using a single Gaussian for the reference pulses at $T = 0.4$ \si{\milli\second}. We look for the fastest solutions with $\tilde\Omega_{\text{peak}}$ and $q$ equal to (or below) these values and denote the corresponding duration as $T_{\text{eqv}}$. Such a solution is found at $T_{\text{eqv}} = 0.25$ \si{\milli\second} by the optimization procedures described in Section~\ref{parametrization}. The thick blue line and red dashed line show for comparison the results from using a Gaussian reference pulse at lower durations. (b) The value of the cost function before and after optimization of 3000 seed reference pulses of 0.25 \si{\milli\second} duration. Blue and green colors indicate sorting of the data in  15 bins with 200 seeds in each bin. We define successful solutions as having a cost value below $C_{\text{success}} = 0.8$ (fat dashed line) after optimization. The probability for finding a successful solution is substantially higher for reference pulses with a low initial cost. The insert shows a histogram of cost values before optimization for 1 million seeds. The seeds are generated by randomizing the control parameters, $\{a_n,t_n^0,w_n\}$, in a suitable interval. The low-cost seeds are seen to be rare. However, the generation and evaluation of a new seed is computationally inexpensive compared to an optimization. This suggests a trade-off between the total number of generated reference pulses and the fraction that we choose to optimize. }
\label{peak_and_q_vs_T}
\end{figure*}

% The relative weight of $\tilde\Omega_{\text{peak}}$ and $q$ in the total cost measure strongly influences the solutions we obtain. By tayloring the weights we can control the trade-off between $\tilde\Omega_{\text{peak}}$ and $q$. We found that by defining the total cost measure as
% \begin{equation}
%   \label{cost function}
% % left(     cost = 5.^(17*peak/omega0-18) + 4.^(3*qval - 3);
%   C = 5^{17\tilde\Omega_{\text{peak}}-18} + 4^{3q - 3},
% \end{equation}
% we obtain solutions with a satisfactory trade-off.

% In addition, a heavy penalty is added if the physical pulses extend beyond the protocol region.
\subsection{Parametrization of a family of reference pulses }
\label{parametrization}

 As any two functions fulfilling the boundary conditions, Eq. (\ref{boundaries}), can be chosen for the reference pulses, it is difficult to search the entire $H_0$-space. In Ref. \cite{du_experimental_2016} the authors parameterize the reference pulses as partially overlapping Gaussians. We hence restrict the search to smooth and time symmetric solutions $\Omega_P(t) = \Omega_S(T-t)$. We choose to define
\begin{align}
  \label{omega S and P}
  \Omega_P(t) &= \Omega_0f(t-T/2-T/10),\\ \Omega_S(t) &= \Omega_0f(-t+T/2-T/10).
\end{align}
The parametrization in Ref. \cite{du_experimental_2016} is extended by choosing the parametrization function, $f(t)$, as a sum of Gaussians,
\begin{equation}
  \label{f}
  f(t) = A \left( e^{-t^2/(T/6)^2 } + \sum_{n=1}^N a_n e^{-(t-t^0_n)^2/(w_n)^2 }\right),
\end{equation}
where $A$ is chosen such that $\max\{f(t)\} = 1$. Here, the amplitude, offsets and widths ($\{a_n,t_n^0,w_n\}$) are the control parameters. We found $N=4$ to offer good solutions. 

At this point the problem is reduced to finding the set of control parameters, $\{a_n,t_n^0,w_n\}$, that minimizes the cost functional, $C$. This is done by locally optimizing several initial guesses, or seeds, for the parameters. The seeds are constructed by choosing random values in a suitable interval. An optimization routine is then employed to iteratively update the control parameters until $C$ is locally minimized, as shown in Fig. \ref{q_and_iter} (b). The optimal physical pulses are then constructed from the optimized control parameters. It is necessary to generate and optimize multiple seeds, since the optimization landscape often contains multiple local optima, as illustrated in Fig. \ref{fig1_H0_and_H_space}.

The \textsc{crab}, \textsc{grape} and Krotov algorithms are widely used as local optimizers on multiple seeds in traditional quantum optimization approaches \cite{crab,grape,tannor,reich_krotov}. However, for these algorithms the cost functional involves the transfer fidelity in addition to the terms from the experimental constraints. This makes the evaluation of the cost functional computationally expensive, as it requires the \sreq{} to be numerically solved. This is not required in our approach, as unit transfer fidelity is guaranteed by the STA formalism. Hence, the cost functional, $C$, can be quickly evaluated and optimized by a standard optimization routine.

\subsection{Results}
The result of the optimization is presented in Fig. \ref{peak_and_q_vs_T} (a). The triangles and squares mark $\tilde\Omega_{\text{peak}}$ and $q$ respectively, obtained by minimizing the cost functional at the given process duration. The dashed-dotted lines mark the target values of $\tilde\Omega_{\text{peak}} = 1.14$ and $q=1.59$, which are the values from Ref. \cite{du_experimental_2016} obtained using Gaussian reference pulses, $f(t) = \exp \left(-t^2/(T/6)^2 \right)$, with $T = 0.4$ \si{\milli\second}. We search for the lowest process duration where we can find equivalent values of $\tilde\Omega_{\text{peak}}$ and $q$. Both $\tilde\Omega_{\text{peak}}$ and $q$ decrease as the process duration increases and the lowest duration where our target is met is found to be $T_{\text{eqv}} = 0.25$ \si{\milli\second}. That is, we find solutions that are nearly twice as fast compared to Ref. \cite{du_experimental_2016} without compromising energy consumption or robustness. The thick and thick-dashed lines show $\tilde\Omega_{\text{peak}}$ and $q$ obtained using Gaussian reference pulses for process durations lower than $T=0.4$ \si{\milli\second}. The optimized pulses perform significantly better at all durations compared to the Gaussian pulse.

% We optimized at process durations from 0.1 \si{\milli\second} to 0.4 \si{\milli\second} in intervals of 0.05 \si{\milli\second}. The result is shown in Fig. \ref{peak_and_q_vs_T} (a). Here we plot $\tilde\Omega_{\text{peak}}$ and $q$ against process duration, $T$. We mark the process duration where we match the values of Ref. \cite{du_experimental_2016} as $T_{\text{eqv}} = 0.25$ \si{\milli\second}. That is, we find solutions that are 60\% faster while maintaining energy requirements and robustness. $\clubsuit$ This is not clear ?  $\clubsuit$

In Fig. \ref{peak_and_q_vs_T} (b) we plot cost values before and after optimization for 3000 seeds. The figure shows that seeds with low cost yield good results more often than seeds with high cost. A seed is taken to be successful if the cost after optimization is below the criterion value, $\log C_{\text{success}} = 0.8$. The probability for finding a successful seed for each bin suggests that only seeds with low cost should be optimized. Such low-cost seeds are rare (see insert in Fig. \ref{peak_and_q_vs_T} (b). However, generating and evaluating a seed is computationally inexpensive, and to produce the data points in Fig.\ref{peak_and_q_vs_T} (a)  we chose to generate 5 million seeds and optimized only the 1000 of them with lowest cost. % The result is shown in Fig. \ref{peak_and_q_vs_T}. We look for the lowest protocol duration where the peak Rabi frequency and sensitivity is equivalent to that of the Gaussian solution at $T = 0.4$ \si{\milli\second}. This is found at $T_{\text{eqv.}} = 0.25$ \si{\milli\second}. That is, we can speed up the population transfer by $60$\% compared to a single Gaussian without sacrificing robustness or energy resources!

% The method can easily be extended to include additional cost measures. The formalism which gives the sensitivity towards Rabi frequency scaling, Eq. (\ref{q}), applies to any kind of pertubation and to any order\cite{ndong_robust_2015}.

% By increasing the complexity of the reference pulse we also increase the complexity of the physical pulse. However, we find that the solutions remain smooth and varies suffeciently slowly, such that they can be implemented by using acousto-optic modulators, as in Ref. \cite{du_experimental_2016}. For example, see the pulses in Fig. \ref{fig1_H0_and_H_space}.

\section{Conclusion}
\label{Conclusion}
We have proposed an optimization strategy for solving state-to-state quantum control problems. Our strategy combines the shortcut-to-adiabaticity formalism and minimization of a cost functional incorporating resource requirements and a perturbative expression for the robustness. Unlike traditional quantum optimal control algorithms our cost functional does not include the transfer fidelity and is therefore computationally inexpensive to evaluate and optimize. We have demonstrated the capability of our strategy on a control problem in the three-level $\Lambda$-system. Here we find solutions that are almost twice as fast as the solution reported in Ref. \cite{du_experimental_2016}, while still obeying experimental constraints. The calculation leading to the perturbative expression for the robustness, Eq. (\ref{fidelity}), can be carried out for any perturbation, and, in principle, to any order \cite{ndong_robust_2015}. This makes our approach especially well-suited for finding solutions that are robust against perturbations. Our strategy can be applied in systems where STA is applicable, including two- and three-level systems \cite{Zhou2016,ruschhaupt_optimally_2012,ruschhaupt_shortcuts_2014,du_experimental_2016}, atoms in harmonic traps \cite{chen_optimal_2011,chen_lewis-riesenfeld_2011,guery-odelin_transport_2014}, quantum many-body systems \cite{STAmanybody,adiabaticTracingManyBody} and quantum heat engines \cite{Campo2014,BoostingHeatEnginePerformance,scalingUpQuantumHeatEngines}. We believe that our approach is a valuable addition to the arsenal of quantum optimal control algorithms, and especially for control problems that require robust solutions.

\section{Acknowledgement}

This work was supported by the European Research Council, the Lundbeck Foundation and the Villum Foundation.

% We have proposed an optimization strategy for solving state-to-state control problems. In this approach we choose a state trajectory connecting the initial and final states under adiabatic conditions and we obtain the control Hamiltonian achieving the same goal in finite time from the STA formalism. The process is then evaluated by a cost functional, representing experimental constraints. We optimize the state trajectory such that the cost functional is minimized, thereby obtaining the time dependent physical Hamiltonian most suitable for experimental realization. We have applied the strategy to a population transfer problem in the three-level $\Lambda$-system, where find solutions that are 60\% faster than the solution reported in Ref. \cite{du_experimental_2016}, while still satisfying the experimental constraints.

% $\clubsuit$ KLAUS: We need to conclude more of offer a wider perspective of the work Any ideas ?  $\clubsuit$

 % For a given a state trajectory, the corresponding control is provided by the STA formalism. In a control theory fashion, we define a cost functional measuring the quality of the control. We then optimize the trajectory such that the cost functional is minimized. We applied the optimization scheme to the population transfer in a three-level $\Lambda$-scheme control problem, and found solutions that are 60\% faster than that of Ref. \cite{du_experimental_2016}, while maintaining the same peak Rabi frequency and sensitivity towards a scaling of the Rabi frequencies.

\bibliographystyle{unsrt}
\bibliography{references}
\end{document}